\documentstyle[preprint,aps]{revtex}
\tightenlines
\def\lsim{\mathrel{\rlap{\lower4pt\hbox{\hskip1pt$\sim$}}
    \raise1pt\hbox{$<$}}}         
\def\gsim{\mathrel{\rlap{\lower4pt\hbox{\hskip1pt$\sim$}}
    \raise1pt\hbox{$>$}}}         
\def\mapright#1{\smash{
\mathop{\to}\limits_{#1}}}

\begin{document}
\draft
\title{SOLAR AND SUPERNOVA CONSTRAINTS ON 
COSMOLOGICALLY INTERESTING NEUTRINOS}
\author{W. C. HAXTON}
\address{Institute for Nuclear Theory, Box 351550 \\ 
and Department of Physics, Box 351560\\
University of Washington,
Seattle, Washington  98195 USA}

\maketitle
\begin{abstract}
The sun and core-collapse supernovae produce neutrino spectra that are 
sensitive to the effects of masses and mixing.  Current results from solar 
neutrino experiments provide perhaps our best evidence for such new neutrino 
physics, beyond the standard electroweak model.  I discuss this evidence as 
well as the limited possibilities for more conventional explanations.  If the 
resolution of the solar neutrino problem is $\nu_e \to \nu_\mu$ oscillations, 
standard seesaw estimates of $m_{\nu_\tau}$ suggest a cosmologically 
interesting third-generation neutrino.  I discuss recent nucleosynthesis 
arguments that lead to an important constraint on this possibility.
\end{abstract}
\pagebreak

\section{Introduction}

The two special lectures presented at Erice under somewhat different titles 
are combined here under the unifying theme I stressed at the school:  What can
 we learn about neutrino masses and mixing from solar and Type II supernova 
studies, and what constraints follow for massive neutrinos in cosmology and 
large-scale structure?  The first half of this paper contains a brief 
review of the current status of the solar neutrino problem and of the evidence 
for matter-enhanced neutrino mixing.  I argue that the alternative of a 
nonstandard solar model is now limited to a single but very interesting class 
of solutions where the core mixes on timescales comparable to those for pp 
chain $^3$He equilibration.  Oscillation of the $\nu_e$ with the $\nu_\mu$ is 
a particularly intriguing solution to the solar neutrino problem, as seesaw 
estimates of the $m_{\nu_\tau}$ then place it near or within the range of 
cosmologically interesting values.  However, this possibility may 
be quite constrained due to growing evidence that r-process nucleosynthesis 
occurs deep within a Type II supernova.  Massive tauon neutrinos with even 
modest mixing angles can destroy the conditions necessary for the r-process.  
This leads to a somewhat distressing situation where massive tauon neutrinos, 
if helpful cosmologically, may have properties that render direct detection 
particularly difficult.  Thus the prospect of continuing hot dark matter 
uncertainties in large scale structure simulations may  be quite real.

One of the experimental pillars of modern cosmology is the nucleosynthesis of 
the light elements within the first few minutes of the big bang.  A second 
reason, in addition to $m_{\nu_\tau}$, for sharing the r-process story with 
the students at this school is to stress its similarities to big bang 
nucleosynthesis.  Each process involves expanding, radiation-dominated nucleon 
gases, nuclear freezeout, and the complications of having to deduce initial 
conditions from a ``fossil" record of abundances.  But the dividends that 
can follow from a detailed understanding of the underlying mircrophpysics is, 
in each case, most significant: from He synthesis we learn about the 
baryon/photon ratio $\eta$ and the number of light neutrino generations; from 
the r-process we may learn about the mass and mixing angles of the $\nu_\tau$.

\section{Solar Neutrinos: Where Are We?}

The original motivation for measuring solar neutrinos was the opportunity to 
make a quantitative check on our theories of stellar evolution and nuclear 
energy generation.  The neutrino fluxes produced by our sun can be measured 
and compared to the predictions of the standard solar model (SSM) [1,2]. 
 This model traces solar evolution 
from the onset of the main sequence, assuming hydrostatic equalibrium (local 
balance between the gravitational force and the gas pressure gradient); energy 
transport by radiation (interior) and convection (outer envelope); and solar 
energy generation by fusion.  The input microphysics includes the opacity as a 
function of temperature and composition, and the nuclear reaction rates for 
the dominant pp chain and CNO cycle conversion of four protons into $^4$He.  
As nuclear reaction energies in the sun are typically $\sim$ 10 keV, the 
nuclear S-factors have to be deduced from somewhat higher energy terrestrial 
cross section measurements by extrapolation.  This requires theory to provide 
the shape S(E) of these nonresonant reactions and to correct for the effects 
of atomic screening in terrestrial targets.  Finally, the model must satisfy 
the boundary conditions of the present day sun (mass, luminosity, radius); 
the solar age ($\sim$ 4.6 Gy); and initial metallicity (abundances for A $>$ 4 
are equated to today's surface abundances, under the assumption that the 
surface has been undisturbed).
  
The resulting sun is a dynamic one with a long-term luminosity rise of $\sim$ 
44\%, a $^8$B neutrino flux that has been significant only in the last 10$^9$ 
years $(\phi(^8$B) $\sim \phi_o e^{\tau/\tau_o}, \tau_o \sim$ 0.9 Gy), and interesting 
nuclear burning scales, such as the time for reaching $^3$He equilibration in 
the pp chain.  Notable successes of the standard solar model include the 
correspondence between the predicted $^4$He abundance near the solar surface 
and the value derived from helioseismology, and the correct depth of the 
convective zone (compare to the value deduced from p-mode oscillations) [1]. 
 Among 
the SSM simplifying assumptions are its one-dimensional form, which allows no 
mixing even during the sun's early convective phase.  One of the often noted 
shortcomings of the SSM, its failure to explain the observed Li depletion by a 
factor $\sim$ 100, is presumably associated with this assumption.

SSM helium synthesis occurs $\sim$ 98\% of the time through the pp chain, 
illustrated in Fig.~1.  The chain is comprised of three cycles (ppI, ppII, 
ppIII) corresponding to three distinct terminations of the fusion.  The ppII 
and ppIII cycles are ``tagged" by associated neutrinos, those from $^7$Be 
electron capture and $^8$B $\beta$-decay, respectively, while the overall rate 
of hydrogen burning is given by the rate of pp and pep neutrinos.  The 
competition between these three cycles is a detailed test of conditions $\--$ 
temperature and composition $\--$ within the solar core.  Thus the detection of 
the associated neutrinos $\--$ measuring the fluxes and spectra illustrated in 
Fig.~2 $\--$ provides an important check on the SSM.

The heroic effort to measure these fluxes began with the $^{37}$Cl detector 
[3] and 
continued with Kamiokande II/III [4] and Superkamiokande [5] and with the 
SAGE/GALLEX 
gallium experiments [6,7].  It continues with the mounting of the Sudbury Neutrino 
Observatory [8], which will begin operations early in 1998.  A fit of the fluxes to 
existing results yields a surprising pattern
\begin{eqnarray}
\phi({\mathrm {pp}})  &\sim& \phi_{\mathrm {SSM}}({\mathrm {pp}}) \nonumber \\
\phi (^7{\mathrm {Be}}) &\sim& 0 \\
\phi(^8{\mathrm B}) &\sim& 0.4 \phi_{\mathrm {SSM}} (^8{\mathrm B}),
\nonumber
\end{eqnarray}
where the subscripts SSM denote SSM values.  These departures from expected 
values lie far outside the known ranges of SSM uncertainties.  (See Fig.~4 of 
Ref.[9]).

There have been many attempts to modify the SSM in order to improve the 
agreement with the results in Eq. (1).  Such ``nonstandard" SMs include ad hoc 
changes in SSM parameters far outside their accepted uncertainties (e.g., 
increasing S$_{11}$, the S-factor for the p+p reaction) as well as new physics 
assumptions (e.g., mechanisms resulting in a significantly reduced heavy 
element abundance in the solar core).  The results of a series of such 
calculations are shown in Fig.~3, taken from Hata et al. [10].  Systematically 
one finds that $\phi(^7$Be) can be suppressed only at the cost of an even 
larger suppression of $\phi(^8$B), in contrast to Eq. (1).  That is, the 
trajectory of $\phi(^8$B) - $\phi(^7$Be) fluxes in Fig.~3 follows a path below 
diagonal, while the experimental results are above the diagonal (Another nice 
illustration of this is found in Castellani et al. [11].)

This difficulty is due to the dependence of the ppII and ppIII cycles on the 
solar core temperature $T_c$.  As $\phi(^8$B) $\sim T^{21}_c$, the observed 
depletion of $\phi(^8$B) in the water Cerenkov experiments requires
\begin{equation}
T_c \sim 0.96 T^{\mathrm{SSM}}_c,
\end{equation}
that is, a cooler core.  But $\phi(^7$Be)/$\phi(^8$B) $\sim T^{-10}_c$, so 
that such a reduced temperature implies
\begin{equation}
{\phi(^7{\mathrm {Be}}) \over \phi (^8{\mathrm B})} \sim 1.5 {\phi^{\mathrm 
{SSM}} (^7{\mathrm {Be}}) \over \phi^{\mathrm {SSM}} (^8 {\mathrm B})},
\end{equation}
in contradiction to Eq. (1).  In other words, the reduced $\phi (^8$B) and 
reduced flux ratio $\phi(^7$Be)/$\phi(^8$B) apparent from Eq. (1) are in 
conflict, with the first requiring a cooler core and the second a hotter 
one.

\section{Matter-enhanced Neutrino Oscillations}

If this argument is completely robust, it appears that current experimental 
results cannot be accommodated by changing the SSM, but instead require new 
particle physics.  While many suggestions have been made, the solution almost 
universely favored, due to its simplicity and plausibility, is neutrino 
oscillations enhanced by matter effects (the Mikheyev-Smirnov-Wolfenstein 
mechanism) [12].

Specializing to the two-flavor case, neutrino oscillations occur if the weak 
interaction eigenstates
\begin{equation} 
|\nu_e \rangle, \, \, | \nu_\mu \rangle
\end{equation}
defined in terms of their accompanying charged leptons, do not correspond to 
the mass eigenstates which diagonalize the free Hamiltonian
\begin{equation}
|\nu_L \rangle, \, \, |\nu_H \rangle
\end{equation}
with mass $m_L$ (light) and $m_H$ (heavy).  Instead there is a nontrivial 
rotation between these two bases, so that
\begin{equation}
|\nu_p (0) \rangle = |\nu_e\rangle = \cos \theta_v | \nu_L \rangle + \sin 
\theta_v|\nu_H \rangle,
\end{equation}
where $|\nu_p(0) \rangle$ is the neutrino of momentum $p$ produced at time t = 
0 by $\beta$-decay.  The particle physics prejudice that $|\nu_e \rangle$ 
should be primarily composed of the light mass eigenstate suggests 
$\theta_v$ is small.  A simple calculation yields downstream of the $\beta$ 
decay source a probability of observing a $\nu_\mu$,
\begin{equation}
|\langle \nu_\mu|\nu_p(t)\rangle|^2 = \sin^2 \, 2\theta_v \sin^2 \, 
\left({\delta m^2 \over 4 E} \, t \right),
\end{equation}
where $\delta m^2 = m^2_H - m^2_L$.  Thus such vacuum oscillations yield 
a small $|\nu_\mu\rangle$ appearance probability proportional to $\sin^2  \, 2 
\theta_v$.

Matter, however, can act as a marvelous regenerator, enhancing this 
oscillation probability because of an adiabatic level crossing.  The neutrino 
index of refraction is modified in matter by charged and neutral current 
interactions, and this effect is flavor dependent because the charge current 
interactions with solar electrons only contributes to the $\nu_e$ forward 
scattering amplitude.  The result is a contribution to the mass matrix 
$M^2$ in the flavor basis of
\begin{equation}
(M^2)_{\nu_e\nu_e} = 4 E \sqrt 2 \, G_F \, \rho_e (x),
\end{equation}
where $E$ is the neutrino energy and $\rho_e(x)$ the local density of 
electrons.  That is, the electron neutrino becomes heavier at high density.

The resulting MSW phenomenon is illustrated in Fig.~4, where $m^2_H (x)$/ 2E 
and $m^2_L (x)/2E$, which are now functions of $x$ because of their dependence 
on $\rho_e(x)$, are plotted relative to their average value.  The relationship 
between the local mass eigenstates $|\nu_L (x)\rangle$ and $|\nu_H (x) 
\rangle$, corresponding to $m_L(x)$ and $m_H(x)$, and $|\nu_e \rangle$ is 
given by Eq. (6), with the important change that $\theta (x)$ now depends on 
$x$.  At $\rho$=0, $\theta (x) = \theta_v \sim 0$, as we have assumed, 
because of our particle physics prejudices, that $|\nu_e \rangle \sim 
|\nu_L\rangle$ in vacuum.  But as Eq. (8) is a positive contribution to 
$(M^2)_{\nu_e\nu_e}$, at sufficiently high density $|\nu_H (x) \rangle 
\rightarrow |\nu_e \rangle$.  That is, as the density increases, $\theta(x)$ 
rotates from $\theta_\nu \sim$ 0 to $\sim \pi/2$.  Furthermore, there is an 
intermediate critical density $\rho(x_c)$ where the matter effects just cancel 
the vacuum mass difference between $(M^2)_{\nu_e\nu_e}$ and 
$(M^2)_{\nu_\mu\nu_\mu}$, leading to the avoided level crossing of Fig.~4.

The transformation to local mass eigenstates leads to a wave equation that is 
diagonal apart from terms depending on d$\rho(x)/dx$.  But if a $|\nu_e 
\rangle$ is produced at $\rho(x_i) > \rho(x_c)$ and $dp(x)/dx$ is everywhere 
ignorable (i.e., dln$(\rho)/dx$ is small over lengths comparable to the inverse 
splittings of the local mass eigenstates of Fig.~4, the propagation is 
adiabatic.
 This corresponds to remaining on the heavy mass trajectory in Fig.~4, 
transforming the 
$|\nu_e\rangle$ into a $|\nu_\mu \rangle$.  Thus nearly complete 
$|\nu_e\rangle \rightarrow |\nu_\mu \rangle$ conversion will occur if [9]:

1)  The initial density $\rho(x_i)$ is sufficient to generate the level 
crossing, $4E\sqrt 2 \, G_F \rho_e (x) \gg \delta m^2$.  

2)  The propagation is adiabatic.  As the separation between mass eigenstates 
is a minimum at $x_c$, this is the point where the oscillation wavelength is 
maximum.  Thus changes in $\rho (x)$ can best be ``seen" at the point.  One 
anticipates, therefore, that the adiabatic condition is most severe at the 
avoided level crossing.

All of this can be worked out analytically using the Landau-Zener trick, as 
described in Ref. [9].  The resulting $\nu_e$ survival probability is
\begingroup
\def\theequation{9a}
\begin{equation}
P^{LZ}_{\nu_e} = {1 \over 2} + {1 \over 2} \cos 2 
\theta_v \, \cos \, 2 \theta_i (1 - 2 P_{\mathrm {hopping}}), 
\end{equation}
\endgroup
where $P_{\mathrm {hopping}}$ is the probability for jumping to the light mass 
trajectory,
\begingroup
\def\theequation{9b}
\begin{equation}
P_{\mathrm {hopping}} = e^{-\pi \, \gamma_c/2},
\end{equation}
\endgroup
and
\begingroup
\def\theequation{9c}
\begin{equation}
\gamma_c = {\sin^2 \, 2\theta_v \over \cos \, 2\theta_v} \, {\delta m^2 
\over 2E} \, {1 \over |{1 \over \rho_c} \, {d \rho (x) \over dx} |_{x = x_c} 
|}.
\end{equation}
\endgroup
Note that $\gamma_c$ depends on the density derivative at the crossing point.  
The adiabatic limit ($P_{\mathrm {hopping}}$ = 0 in Eq. (9a)) was derived by 
Bethe [13], while $P_{\mathrm {hopping}}$ was derived by Haxton [14] and 
independently 
by Parke [15].  The two conditions above correspond to the initial local mixing 
angle $\theta_i \sim \pi/2$ and to $\gamma_c \gg 1$, yielding $P^{LZ}_{\nu_e} 
\sim {1 \over 2} - {1 \over 2} \cos \, 2\theta_v  \sim 0$.

A search for a fit to the experimental results gives the familiar iso-SNU plot
shown in Fig.~5.  The better fit to the data is given by the small mixing 
angle solution of $\sin^2 \, 2\theta_v \sim$ 0.005 and $\delta m^2 \sim 6 
\cdot 10^{-6} eV^2$.  This corresponds to strong conversion of the 
intermediate energy $^7$Be neutrinos; partial depletion of the $^8$B 
neutrinos, strongest at the low energy end; and survival of most of the pp 
neutrinos.  This occurs because the $^7$Be neutrinos have an adiabatic level 
crossing; almost all of the lower energy pp neutrinos do not have a level 
crossing and thus evade conversion; while the higher energy $^8$B neutrinos 
straddle the adiabatic boundary, so that the lower energy portion of the 
spectrum is converted more strongly than the higher energy end.  Thus the MSW 
mechanism has a dramatic signature, a characteristic energy-dependent 
distortion of the $\nu_e$ spectrum and the appearance of $\nu_\mu$ neutrinos.

If this is the solution to the solar neutrino problem, it may have important 
consequences for those working in cosmology and large scale structure.  
Generalizing the above discussion to three mass eigenstates, with $m_1 \ll m_2 
\ll m_3$ and $|\nu_1\rangle \sim |\nu_e\rangle$, then an attractive choice is 
$m_2 \sim \sqrt{m_2^2 - m_1^2} \sim \sqrt{\delta m^2} \sim$ few $\cdot 10^{-3}$ 
eV.  That is, if we suppose the oscillation is $\nu_e \rightarrow \nu_\mu$ 
rather than $\nu_e \rightarrow \nu_\tau$, we can estimate the ``muon" neutrino 
mass.

If one thinks in terms of the more general multiplets that might exist in 
extensions of the standard model, e.g.,
\begingroup
\def\theequation{10}
\begin{equation}
\left(
\begin{array}{c}
  u \\ d \\ \nu_e \\ e^-
\end{array}
\right)
\end{equation}
\endgroup
the much smaller mass of the $\nu_e$ relative to the other first-generation 
fermions is a bit of a puzzle.  One would have assumed that all of these 
particles will have similar couplings to the mass-generating fields, and 
thus might have comparable masses $\sim m_D$.  One nice resolution of this 
problem comes from the observation that $\nu_e$ is unique in not having a 
charge $\--$ or any other additively conserved quantum number.

Thus neutrinos can also have Majorana masses, e.g.,
\begingroup
\def\theequation{11}
\begin{equation}
M_R \overline{\nu^c_R} \nu_R.
\end{equation}
\endgroup
If the right-handed Majorana mass is large, characterizing the scale of some 
new physics, the resulting diagonalization of the mass matrix yields one heavy 
eigenstate and one light one
\begingroup
\def\theequation{12}
\begin{equation}
\sim m_D \left({m_D \over M_R}\right).
\end{equation}
\endgroup
This ``seesaw" mechanism [16] thus generates the needed small parameter, 
$m_D/m_R$, 
to explain the lightness of the neutrino relative to its charged partners.
Now Eq. (12) suggests that the mass relations $m_{\nu_e} : m_{\nu_\mu} : 
m_{\nu_\tau}$ might be related to the squares of the Dirac masses for the 
corresponding upper isospin quarks, $m^2_u: m^2_c: m^2_\tau$.  Thus 
$``m_{\nu_\mu}" \sim$ few $\cdot 10^{-3}$ eV allows us to estimate the $M_R$ 
and predict $m_{\nu_\tau} \sim (1 -$ few) eV.  That is, an MSW $\nu_e 
\rightarrow \nu_\mu$ solution to the solar neutrino problem is attractive 
because it allows the $\nu_\tau$ to play a cosmologically interesting role, 
perhaps being a source of hot dark matter helpful in explaining the formation 
of large-scale structure.

\section{Is there no alternative to new particle physics?}

Because the conclusion reached in the last section is a profound one, having 
far-reaching consequences for both particle physics and cosmology, it is 
important to ask whether there is any more conventional ``escape".  Andrew 
Cumming and I recently found that there is one exception to the $T_c$ 
arguments in Eqs. (2) and (3), and thus one possible path to a nonstandard 
model that might accommodate Eq. (1).  Because this work is described in a 
publication and in another talk [17], I will present only a brief summary here.

As illustrated in Fig.~3, many SSM modifications have been tried and found to 
depart from the pattern in Eq. (1).  This suggested to us that if an 
acceptable nonstandard SSM possibility exists, its underlying physics might be 
subtle and thus difficult to anticipate.  Therefore we decided to try a naive 
approach, exploring nonstandard SMs phenomenologically, in the hope that this 
might point the way to the necessary change in the SSM.

As we wanted our explorations to be reasonable, we required that our 
phenomenological adjustments preserve certain properties.  This included 
reproducing the correct luminosity; retaining SSM microphysics, given the work 
that has been invested in determining accurate S-factors and opacities; and 
demanding that the model be steady-state.  The motivation for the last 
condition was to avoid solutions where today is somehow a special time in the 
sun's history.

In the SSM the steady-state condition is satisfied locally: the production of 
various ``catalysts" of the pp chain like d, $^3$He, and $^7$Be is equated to 
the local consumption, once equilibrium is reached.  This is not a physics 
result in the SSM, but a consequence of its assumptions (no mixing in one 
dimension).  Thus the ``models" we explored phenomenologically were 
considerably less constrained, allowing transport of the ``catalysts" 
mentioned above as well as H and $^4$He.  

Our procedures are described in [17] and will not be repeated here.  They led 
us to focus quickly on $^3$He, which offers the prospect of transport on 
interesting time scales.  Equating the production rate of $^3$He through the 
p + p 
reaction to its primary destruction rate, $^3$He + $^3$He, leads to an 
estimate of the SSM equilibrium abundance
\begingroup
\def\theequation{13}
\begin{equation}
X^{eq}_3 \sim 7 \cdot 10^{-4} X_1 \, T^{-6}_7
\end{equation}
\endgroup
and of the time required to reach equilibrium
\begingroup
\def\theequation{14}
\begin{equation}
\tau^{eq}_3 \sim {1 \over X_1} \, T^{-10}_7 
\end{equation}
\endgroup
where $X_1$ is the local abundance of H and $T_7$ the temperature in 10$^7$ K. 
 For example, 99\% of equilibrium is reached at r $\sim 0.1 R_{\odot}$ after 
$\sim 5 \cdot 10^6$ years of SSM burning, and at t = 4.6 Gy equilibrium has 
been achieved for r$\lsim 0.27 R_\odot$.  The resulting $^3$He SSM profile is 
thus characterized by a steep gradient over the energy-producing core, shown in 
Fig. 6.
  
Our exercise showed that a steady-state solar model can produce neutrino 
fluxes reasonably close to Eq. (1) only if the core mixes in a prescribed way 
on a timescale characteristic of $^3$He equilibration, as illustrated in 
Fig.~7.  The circulation is that of ``elevator convection": localized, 
relative rapid downward flow of $^3$He rich material in narrow plumes, followed by a 
slower, broad restoring flow to large r.

It is important to understand on general grounds, why such circulation would 
produce a flux pattern similar to Eq. (1).  First, the competition between the 
ppI cycle and the ppII + ppIII cycles depends on that between $^3$He + 
$^3$He, which is quadratic in $X_3$, and $^3$He + $^4$He, which is linear.  The 
downward flow with the velocity shown in Fig.~7 sweeps $^3$He-rich material 
deep into the core (where $X^{eq}_3$ is quite small and $\tau_3^{eq}$ quite 
short), leading to its ignition.  The resulting out-of-equilibrium $^3$He 
burning enhances the ppI cycle relative to ppII + ppIII in proportion to 
$X_3/X^{eq}_3$ ratio achieved at ignition.  Second, the $^3$He + $^4$He 
reactions that do occur take place at small r and high T, where the ppII/ppIII 
branching ratio $\sim T^{-10}$ favors ppIII over ppII.  Thus the net effect of 
the circulation is to strongly reduce $\phi(^7$Be) and somewhat reduce 
$\phi(^8$B).  If such mixing were to occur in the sun, the core temperature 
would be reduced to about 0.96$T_c^{\mathrm{SSM}}$ because ppI burning is more 
efficient.

The naive $T_c$ argument invoked in Section 2 to argue against nonstandard 
models is therefore not universally valid.  But, in the context of 
steady-state models with standard microphysics, existing neutrino flux results 
seem to leave only one nonstandard model possibility open: core mixing on time 
scales of $^3$He equilibration.
  
What is surprising about this possibility $\--$ given that no solar physics 
went into its deduction $\--$ is that it has some physical plausibility.  The 
possibility of large-scale collective flow induced by the SSM $^3$He gradient 
is a long-standing concern:  even recent explorations of the known SSM ``solar 
spoon" overstability conclude that mixing of some amplitude could result [18].
The continuous mixing envisioned in Fig.~7 is attractive in this regard, because 
there is no composition gradient working against the flow, as the core is 
kept homogeneous in H and $^4$He by the mixing.  Another 
interesting feature is the circulation time of a few $\cdot 10^7$ years.  This 
reminds one of a concern, originally expressed by Roxburgh [19], that the sun's 
early convective core, arising from out-of-equilibrium burning of the CNO 
cycle over the first 10$^8$ years of the main sequence, could then persist 
because of the growth of the $^3$He gradient.

As the arguments Andrew Cumming and I made have been misconstrued by some, I 
would like to ``summarize my points" clearly:

$\bullet$ It is obvious that the MSW mechanism is elegant and has profound 
implications.  It is the solution to the solar neutrino problem I favor.

$\bullet$ However existing generic arguments that no nonstandard model 
modification can lead to the flux pattern of Eq. (1) appear to be overstated.

$\bullet$ In the case of steady state models with standard microphysics, 
existing neutrino flux measurements appear to have narrowed the possibilities 
to a single class, those with core mixing on timescales of $^3$He 
equilibration.  This is remarkable.

$\bullet$ Somewhat surprisingly, this class may also be the one with the most 
physical plausibility.  The core mixing possibilities associated with the 
solar spoon overstability and a persistant convective core have been discussed 
for many years.

$\bullet$ The question of whether such core mixing would arise in a 3D (or 2D) 
solar model is, at this point, purely speculative.  Conversely, the absence 
of mixing in the SSM is not a physics result, but a reflection of its 1D 
character, i.e., of assumptions made at the outset.

$\bullet$ If a viable model could be constructed, the differences with the SSM 
would have consequences for a range of astrophysical issues:  galactic $^3$He 
evolution, the evolution of solar-like stars along the color-magnitude 
diagram, and helioseismology.  The last is likely to be a very 
difficult test for the proposed mixing scheme to satisfy, as the
molecular weight profile in the solar core would be quite different
from the standard solar model.  However I have yet to see a 
published argument on this point that I would regard as definitive.

$\bullet$ The challenge to theory to construct a 3D hydrodynamic model of the 
sun is considerable.  Therefore it is likely that SNO and Superkamiokande will 
prove a solution to the solar neutrino puzzle before theory progresses on this 
issue.

$\bullet$ At one point the ``negative $\phi(^7$Be)" issue seemed to hint that 
the solution had to be particle physics.  The most recent Superkamiokande 
result (306 days), $\phi (^8$B) = 2.44 $\pm 0.06^{+0.05}_{-0.09} \cdot 
10^6/$cm$^2$s, may weaken this claim.  The corresponding $^{37}$Cl result of 
2.55 
$\pm$ 0.25 SNU yields, for $\sigma(^8$B) = (1.11 $\pm 0.05) \cdot 10^{-42}$
cm$^2$, the limit  $\phi(^8$B) $\lsim (2.30 \pm 0.22 \pm 0.05) \cdot 
10^6/$cm$^2$s.  Thus this 
accommodates all of the Superkamiokande range at 1 $\sigma$ without the need 
for a negative $\phi(^7$Be). (Of course, one must make room for a small 
pep/CNO contribution to the $^{37}$Cl experiment; but the somewhat lower 
Superkamiokande result helps reduce the difficulty of achieving this).
This emphasizes how crucial SNO and low-energy 
Superkamiokande data will be to proving neutrino oscillations.

$\bullet$ Again, the physics possibility sketched here is not offered as a 
solution to the solar neutrino puzzle, but as an argument that a nonstandard 
solar model solution remains an open possibility.  It is not my favorite 
solution, nor is it one I'm prepared to completely rule out at this time.

\section{Big-bang Nucleosynthesis vs. the r-process}

Despite the cautions expressed in the previous section, I would like to 
further explore the consequences of a neutrino physics resolution of the solar 
neutrino puzzle.  As discussed in Section 3, a $\nu_e \rightarrow \nu_\mu$ 
explanation of the missing solar neutrinos is nicely compatible with a 
cosmologically interesting $m_{\nu_\tau}$.  However, there is growing evidence 
that the cosmological role of the $\nu_\tau$ is limited by constraints from 
r-process nucleosynthesis.  This is the theme of this second lecture.

Given the cosmological bent of the audience, I would like to begin by drawing 
parallels between big bang nucleosynthesis $\--$ a cornerstone of modern 
cosmology $\--$ and nucleosynthesis in a supernovae.  In the big-bang one 
encounters:

$\bullet$ An expanding, radiation-dominated, proton rich gas.  Below about $T 
\sim$ 1 MeV the weak interactions have frozen out, prior to nucleosynthesis.

$\bullet$ A nuclear freezeout occurs at $T \sim$ 100 keV $\sim 10^9$ K, when 
the n + p $\rightarrow$ d + $\gamma$ bottleneck is broken.  The resulting 
$^4$He/H 
ratio depends on the n/p ratio at freezeout ($\sim$ 1/7).

$\bullet$ The nuclear gas is relatively dilute, so three-body reactions are 
rare.  The absence of stable nuclei at A = 5 and 8 terminates the reaction 
chains, as $\alpha + \alpha$ and $\alpha$ + p then cannot proceed.

$\bullet$ Today a fossil record of the big bang exists in the abundances of H, 
$^3$He/d, $^4$He, and Li.

$\bullet$ The clarity with which this primordial record can be read depends on 
our ability to correct for the effects of subsequent galactic chemical 
evolution.  Figure 8, showing the complicating effects of both stellar 
production and destruction of Li, is one illustration of the challenges.

$\bullet$ Our resulting understanding of big bang nucleosynthesis has yielded 
fundamental constraints on cosmological and particle physics parameters, 
determining the baryon/photon ratio $\eta$ and constraining the number of 
light 
neutrino generations.

The less familiar (to this audience) conditions found near the mass cut of a 
Type II supernova are quite similar, yet the few differences have interesting 
consequences:

$\bullet$ This ``hot bubble" region in a supernova is an expanding, radiation 
dominated, neutron rich gas.  Neutrinos streaming through this region are 
thermally decoupled from the matter.

$\bullet$ Important nucleosynthesis occurs as this material expands off the 
protoneutron star, dropping in temperature from T $\sim$ 300 keV to T $\sim$ 
100 keV, where reactions freeze out.  The n/p ratio is crucial to this 
synthesis.

$\bullet$ The nucleon gas is sufficiently dense that reactions such as 3 
$\alpha \rightarrow ^{12}$C can bridge the mass gap at A = 8.  The $\alpha$ 
process is thought to proceed up to medium mass  nuclei, producing a gas 
dominated by $^4$He, a few heavy seed nuclei, and excess neutrons.

$\bullet$ The resulting rapid capture of neutrons on the seeds produces the 
r-process nuclei.  The current abundances of these nuclei constitute a fossil 
record of past galactic supernova nucleosynthesis.

$\bullet$ An understanding of this synthesis can yield important constraints on 
particle physics and astrophysics parameters, such as the mass and mixing of 
the $\nu_\tau$ and the supernova rate averaged over the galaxy's lifetime.
  
The recent development is the convergence of several arguments which place the 
site of the r-process deep within Type II supernova.  I will first describe 
the r-process, then review these arguments.

\section{The r-process and Its Astrophysical Site}

Consider a stellar environment where a neutron gas is present together with 
nuclei. Within the nucleus the surface of the neutron/proton Fermi seas are 
$\sim$ 8 MeV below the continuum, a value presumably much above stellar 
temperatures.  We assume that the neutron capture rate $(n, \gamma)$ is slow 
compared to typical nuclear $\beta$ decay rates.  This then allows the weak 
interaction to maintain the equilibrium of the proton/neutron Fermi seas as 
neutrons are captured.  Thus the neutron-induced nucleosynthesis proceeds 
along a path in (N,Z) centered on the stable nuclei.  This is called the s- or 
slow-process.  The rate of nucleosynthesis, that is, the rate of change of A = 
N + Z, is then controlled by the neutron capture rate.

The r- or rapid-process requires more exotic stellar conditions.  The neutron 
capture rate is fast compared to $\beta$ decay and thus determines a new 
equilibrium condition:  quantum levels above the usual neutron Fermi sea in 
the nucleus fill to within a distance $\propto$ T of the continuum, where T is 
temperature of the neutron gas, as (n, $\gamma) \leftrightarrow (\gamma$ n) 
comes into balance.  The nucleosynthesis rate is then proportional to the 
$\beta$ decay rate:  any beta decay of a neutron to a proton opens up a hole in the 
neutron sea, which then is rapidly refilled.  The nucleosynthesis path is 
along very exotic neutron-rich nuclei, determined by the (n$, 
\gamma)\rightarrow(\gamma$, n) equilibrium.  If a $\beta$ decay rate for a 
particular nucleus (Z,N) is slow, the mass flow is restricted at the point, 
increasing the abundance.  Thus the abundance for (Z,N) is expected to be 
inversely proportional to the $\beta$ decay rate at that Z and N.

Nuclei exhibit gaps in their level structure at closed shells, e.g., N $\sim$ 
82 and $\sim$ 126.  When such a gap is encountered, the mass flow, controlled 
by $\beta$ decay, is redirected along N = constant in the (Z,N) plane, until 
the gain in (Z,N) symmetry energy is sufficient to bridge the gap, bringing 
the next neutron level below the continuum.  
As $\beta$ decay rates near closed shells are low, 
the closed neutron shells produce large abundance peaks, as can be seen in 
Figs. 9 and 10.

Once the neutron exposure ends, the r-process products decay back to the valley 
of stability, (Z,N) $\rightarrow$ (Z + $\Delta, N - \Delta$) by repeated 
$\beta$ decay.  Spallation following $\beta$ decay can shift A to somewhat 
lower values.

About half of all nuclei above A $\gsim$ 80, including all the transuranics,
are synthesized by the r-process.  For example the s-process chain

$$
(Z,N)^s_{ee} + n \rightarrow (Z,N + 1)_{eo} \mapright{\beta}\
 (Z+1, N)^s_{oe} + n
\rightarrow
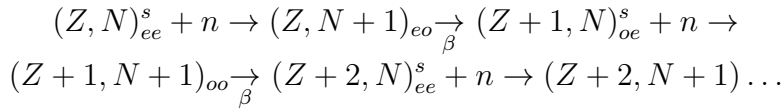
$$
$$
(Z+1, N +1)_{oo} \mapright{\beta}\
 (Z + 2, N)^s_{ee} + n \rightarrow
(Z + 2, N + 1) \dots 
$$
flows through A = Z + N +1 and Z + N + 2, producing the stable odd-even and 
even-even isotopes $(Z + 1, N)^s_{oe}$ and  $(Z + 2, N)^s_{ee}$, but bypasses 
the even-even nucleus (Z, N + 2) which frequently is stable.  But this 
neutron-rich bypassed isotope would be produced by the r-process.  Thus 
some nuclei are uniquely due to the r-process, while others can be synthesized 
by both the s- and r-processes or only by the s-process.

It has been known for almost four decades that the r-process requires 
spectacularly explosive conditions
\begingroup
\def\theequation{15}
\begin{eqnarray}
\rho_n &\sim& 10^{20}/{\mathrm {cm}}^3 \nonumber \\
T &\sim& (1-3)\cdot 10^9 K \\
t &\sim& 1 \,\, {\mathrm {sec}} \nonumber
\end{eqnarray}
\endgroup
where $\rho_n$ is the neutron density.
Suggested primary sites $\--$ those requiring no pre-enrichment of s-process 
elements to serve as ``seeds" for the neutron capture $\--$ include the 
neutronized atmospheres above the protoneutron stars in Type II supernova 
explosions, neutron-rich jets from supernovae or neutron star mergers, and 
inhomogeneous big bangs.  Secondary sites $\--$ those with pre-existing seeds 
$\--$ can support successful r-processes with somewhat lower $\rho_n$.  
Suggestions have included the He and C zones in Type II supernovae and the red 
giant He flash.

There is a growing body of evidence favoring a primary r-process in Type II 
supernova.  The discovery of very metal poor halo stars, [Fe/H] $\sim$ -1.7 
and - 3.12, enriched in r-process material, argues for a primary process, 
already operating in the early history of the galaxy [21].  Studies of galactic 
chemical evolution [22] have found that the growth of r-process material is 
consistent with low-mass Type II supernovae being the r-process site.  
Finally, the suggestion made long ago that the r-process might be associated 
with the expansion and cooling of neutron-rich matter from the vicinity of the 
mass cut in supernovae [23] has been modeled much more convincingly.  It has 
been shown in Ref. [24] that an expanding neutron-rich nucleon gas can undergo 
an $\alpha$-particle freezeout, on which effectively all of the protons are 
consumed, followed by an $\alpha$-process, in which seed nuclei near A = 100 
are produced. The r-process then takes place through the capture of excess 
neutrons on these seeds.  While this specific model has some shortcomings 
$\--$ overproduction of $^{88}$Sr, $^{89}$Y, and $^{90}$Zr and the need for 
very high entropies $\--$ it has demonstrated that a supernova ``hot bubble" 
r-process can produce both a reasonable abundance distribution and an 
appropriate amount of r-process ejecta.

\section{Supernova Neutrinos and the r-process}

In the infall stage of a core collapse supernova, neutrinos are trapped once a 
density of $\sim$ 10$^{12}$g/cm$^3$ is reached, guaranteeing that the 3 $\cdot 
10^{53}$ ergs of released gravitational energy is locked within the core 
until after core bounce.  Eventually 99\% of this energy is released after 
core bounce in the form of neutrinos, which diffuse outward to the trapping 
radius (or neutrinosphere) on a time scale $\tau \sim$ 3 sec.  During this random 
walk the neutrinos remain in weak equilibrium through interactions of the type 
$\nu_e \overline{\nu}_e \leftrightarrow \nu_\mu \overline{\nu}_\mu$
which guarantees approximate equipartition of the energy per 
flavor.  However the final decoupling of the neutrinos from the matter is 
flavor dependent due to the stronger $\nu_e + e^- \leftrightarrow \nu_e + e^-$ 
cross section and the charge current reactions
\begingroup
\def\theequation{16a}
\begin{equation}
\nu_e + n \leftrightarrow p + e^- 
\end{equation}
\endgroup
\begingroup
\def\theequation{16b}
\begin{equation}
\bar \nu_e + p \leftrightarrow n + e^+.
\end{equation}
\endgroup
The net result is a characteristic hierarchy of temperatures
\begingroup
\def\theequation{17a}
\begin{equation}
T_{\nu_\mu, \bar \nu_\mu, \nu_\tau, \bar \nu_\tau} \sim 8 \,\,{\mathrm {MeV}} 
\end{equation}
\endgroup
\begingroup
\def\theequation{17b}
\begin{equation}
T_{\bar \nu_e} \sim 5 \,\,{\mathrm {MeV}} 
\end{equation}
\endgroup
\begingroup
\def\theequation{17c}
\begin{equation}
T_{\nu_e} \sim 4 \,\, {\mathrm {MeV}}\,,
\end{equation}
\endgroup
where the $\bar \nu_e/\nu_e$ difference is a result of the neutron richness of 
the matter, which enhances the reaction in (16a) and keeps the $\nu_e$ coupled 
until it reaches a somewhat larger radius and correspondingly lower T.

Due to work by Woosley, Haxton, et al. [25] and by Domagatskii and 
collaborators [26], 
it has been appreciated that neutrinos, on passing through the mantle of the 
supernova, can be responsible for novel nucleosynthesis.  A much discussed 
example is the production of $^{19}$F by $(\nu, \nu')$ spallation within the 
Ne zone.  This is an effective method of synthesizing the $^{19}$F found in 
our galaxy because the $^{19}$F/$^{20}$Ne ratio is small, $\sim$ 0.0003.

Now the neutrino process production of $^{19}$F occurs in the Ne shell, (2-3) 
$\cdot 10^4$ km from the protoneutron star.  The ``hot bubble" r-process, 
though occurring at $\sim$ 10 sec after core bounce when the majority of 
neutrinos have already escaped, takes place at 600-1000 km, where the neutrino 
flux is far more intense.  Thus the question that occurred to my collaborators 
and me:  Could there be a ``neutrino finger print" on the r-process 
distribution that would prove it occurred in an intense neutrino flux?

It is clear that neutrino effects during the r-process may be ``erased" 
because A ($\nu, \nu')A^*$ reactions are masked by the much stronger effects 
of photoabsorption.  Yet this leaves open the possibility of interesting 
postprocessing effects, occurring after the r-process has frozen out at T 
$\sim 10^9$ K.  To assess these effects, we integrated backward in time, 
subtracting from the observed r-process distribution the effects of neutrino 
spallation following freezeout.  This ``inversion" produces, from the final 
r-process distribution found in nature, the ``true" 
r-process distribution at freezeout as a function of the assumed neutrino 
fluence after freezeout.

The results [27] are given in Figs. 10 and 11.  The neutrinos have a general 
smoothing effect and dramatically change the distribution in two places, the 
``valleys" below the abundance peaks at A $\sim$ 130 and $\sim$ 195.  This is 
easily understood:  the effects of neutrino postprocessing, where neutrino
reactions transfer enough energy to typically knock 3-5 neutrons out of the neutron-rich 
target nucleus, are magnified in this 
region because the parent nuclei (in the abundance peaks) are very abundant.  The analogy with $^{19}$F 
and $^{20}$Ne is clear.

In particular, eight nuclei, lying in the windows A = 124 - 126 and 183 - 187, are 
inordinately sensitive to postprocessing.  Initially we thought that, by 
demanding these nuclei not be overproduced by the postprocessing, we could 
only place an upper bound on the neutrino fluence following freezeout.  These 
window isotopes are not produced at a significant level in most r-process 
calculations.  We found, instead, that their abundances could be fit at $\sim 
1 \sigma$ under the hypothesis of neutrino-induced nucleosynthesis, assuming a 
fluence slightly below our limiting value.  The results are
$$
{\cal F} (A = 124 - 126) \sim 0.031 
$$
$$
{\cal F} (A = 183 - 187) \sim 0.015 
$$
where the fluences are per flavor and specified in units of 10$^{51}$ 
ergs/(100 km)$^2$.  These values correspond closely to those of the ``hot 
bubble" r-process [24]. The lower value of ${\cal F}$ for A = 183 - 187 is natural, 
corresponding to later synthesis of the A $\sim$ 195 mass peak.  Thus a 
``neutrino finger print" of the ``hot bubble" r-process is found, providing 
another strong argument for a primary Type II supernova r-process.

\section{Oscillations of Tauon Neutrinos}

In Section 3 I argued that $\nu_e \rightarrow \nu_\mu$ oscillations are an 
attractive explanation of the solar neutrino problem, leaving a massive 
$\nu_\tau$ free to play an important role in cosmology as hot dark matter.  
But, extending Fig. 4 as shown in Fig. 11, this then leads to a second, $\nu_e 
\leftrightarrow \nu_\tau$ crossing at higher densities.  For $m_{\nu_\tau}$ in 
the cosmologically interesting window of (1-50) eV, this crossing occurs 
outside the neutinosphere in a supernova, i.e., after the neutrinos have 
decoupled from the matter.  The result is an interchange in the corresponding 
spectra, producing anomalously hot $\nu_e$s,
$$
T^{\mathrm {MSW}}_{\nu_e} \sim 8 {\mathrm {MeV}} \gg T_{\bar \nu_e}, 
$$
The $\bar \nu_e$ experiences no crossing.

Now the cross sections in Eqs. (16) vary as T$^2$, while the luminosity is 
approximately equipartitioned in flavor.  Consequently the rate for $\nu_e + n 
\rightarrow e^- + p$ in the hot bubble is approximately doubled if $\nu_e 
\rightarrow \nu_\tau$ oscillations occur, while the $\bar \nu_e + p 
\rightarrow e^+ + n$ rate remains fixed.  Thus the $\nu_e \rightarrow 
\nu_\tau$ crossing can drive the ``hot bubble" proton rich, destroying any 
prospect for an r-process.

The necessary conditions for an adiabatic $\nu_e \leftrightarrow \nu_\tau$ 
crossing affecting the r-process have been given by Fuller and Qian [28].
 The 
existence of a ``hot bubble" r-process puts stringent limits on a 
cosmologically interesting $\nu_\tau$.

To make the discussion specific, let's suppose a 7 eV neutrino is advocated by 
large-scale structure studies.  What are the consequences?

$\bullet$ Such a m$_{\nu_\tau}$ is acceptable for modest $\theta_{e\tau} \gsim 
10^{-2}$ only if the above r-process argument is wrong.  (That is, the 
r-process actually occurs elsewhere). I think this is unlikely.  More 
important, this can be checked in terrestrial detectors by measuring T$_{\nu_e}$ and T$_{\bar \nu_e}$ 
for the neutrinos emitted by the next galactic supernova, an event we should 
not miss!

$\bullet$ The m$_{\nu_\tau}$ is $\sim$ 7 eV and the r-process argument is 
evaded by having a small mixing angle, $\theta_{e \tau} \lsim$ 0.003.  This 
would mean that a $\nu_e \leftrightarrow \nu_\tau$ oscillation fails to occur 
in a supernova because the small mixing angle leads to a nonadiabatic 
crossing.

Now this is not implausible since such a $\theta_{e \tau}$ is consistent with 
possible ranges for third-generation CKM angles
$$
0.002 < V_{ub} < 0.007 
$$
$$
0.003 < V_{t d} < 0.018,
$$
and thus fits with some (probably poorly justified) prejudices.  But this has 
the ugly consequence that terrestrial $\nu_e \rightarrow \nu_\tau$ 
disappearance experiments $(\sin^2 \, 2 \theta_{e \tau} \lsim 4 \cdot 10^{-5})$ 
are 
then very difficult.  The best hope to experimentally confirm a hypothesized 
hot dark matter $\nu_\tau$ would then be a more favorable 
$\theta_{\mu\tau}$, leading to a measurable $\nu_\mu \rightarrow 
\nu_\tau$ disappearance. 
  
Of course, it could turn out the $\theta_{\mu\tau}$ is also quite small, 
frustrating oscillation searches in this channel, too.  It then 
might prove difficult to demonstrate the existence of a cosmological 
interesting $\nu_\tau$ in the laboratory, a prospect large scale structure 
theorists might find annoying.

$\bullet$ Finally there is one other way out:  the mixing angle could be large 
and the r-process argument correct, yet r-process consequences could be evaded 
by making $m_{\nu_\tau} \lsim$ 3 eV. The necessary 7 eV of dark matter could be 
achieved with $m_{\tau_\nu} \sim m_{\nu_\mu} \sim m_{\nu_e}$.  This scenario 
has been recently invoked in order to make sense out of a variety of 
astrophysical and terrestrial hints of massive neutrinos [29].

The good news is that SNO and Superkamiokande should help us take the first 
step, deciding whether massive neutrinos are responsible for the solar 
neutrino problem.  This will be a great help in reading the ``tea leaves" of a 
cosmologically interesting $\nu_\tau$.

I thank the Ettore Majorana International Centre for hosting this school and 
Prof. Norma Sanchez for her invitation and able leadership of the school.  
The research reported here was supported by the US Department of Energy.

\pagebreak

\begin{figure}
\caption{The solar pp chain.  Note that each of the three 
cycles, ppI, ppII, and ppIII, can be associated with a distinct
neutrino.}
\end{figure}

\begin{figure}
\caption{The flux densities (solid lines) of the principal $\beta$ decay 
sources of solar neutrinos of the standard solar model.  The total fluxes are 
those of the SSM of Ref. [1].  The $^7$Be and pep electron capture neutrino fluxes 
(dashed lines) are given in units of cm$^{-2}$ s$^{-1}$.}
\end{figure}
  
\begin{figure}
\caption{The fluxes allowed by the combined results of the Homestake, 
SAGE/GALLEX, and Kamiokande experiments compared to the uncertainties of the 
SSM (upper ellipse) and to various nonstandard model predictions.  The solid 
line is the T$_c$ power law (From Ref. [10]).}
\end{figure}
  
\begin{figure}
\caption{Schematic illustration of the MSW level crossing.  The dashed lines 
correspond to the electron-electron and muon-muon elements of the mass matrix. 
Their intersection defines the level crossing density $\rho_c$.  The solid 
lines are the trajectories of the local heavy and light mass eigenstates.  If 
an electron neutrino with a suitable vacuum mass is produced deep in the solar 
core and propagates adiabatically, it will follow the heavy mass trajectory, 
emerging from the sun as a $\nu_\mu$.}
\end{figure}
  
\begin{figure}
\caption{The MSW solutions allowed at 95\% confidence level, given the 
standard model fluxes of Ref. [1].  From Ref. [10].}
\end{figure}
  
\begin{figure}
\caption{The dashed line represents the SSM $^3$He profile after $\sim$ 4.6 Gy 
of burning.  The solid line indicates where $^3$He would be burned given the 
convection pattern of Fig. 7.}
\end{figure}
  
\begin{figure}
\caption{The convection pattern required to suppress both $\phi(^8$B) and 
$\phi(^7$Be)/$\phi(^7$B).  The downward flow is in plumes, rapid and 
localized, requiring $\sim$ few $\cdot 10^6$ years.  This leads to 
out-of-equilibrium burning of $^3$He at small r.  The slow, broad upward flow 
allows the cycle to replenish the $^3$He.  Typical upward times are $\sim$ 
few 10$^7$ years.}
\end{figure}
  
\begin{figure}
\caption{Li abundance in solar-like stars as a function of metallicity [Fe/H], 
from Timmes et al. [20].  The region between the dashed curves represent the 
range of neutrino process contributions.}
\end{figure}
  
\begin{figure}
\caption{The solid curve is the r-process production that, when combined with 
neutrino postprocessing, would produce the observed abundances (dashed line). 
It is assumed that the production in the window A = 124 - 126 is entirely due 
to postprocessing.}
\end{figure}
  
\begin{figure}
\caption{As in Fig. 9, only for the A $\sim$ 195 mass peak.
Again, the production in the window 183-187 is assumed to be due
only to neutrino postprocessing.}
\end{figure}
  
\begin{figure}
\caption{The three-flavor level crossing diagram, analogous to Fig. 4,
showing that a second crossing may occur at higher densities.}

\end{figure}
  
\end{document}